\documentclass[onecolumn]{aastex631}

\usepackage{multirow}
\usepackage{amsmath}
\usepackage{appendix}
\usepackage{CJK}

\shorttitle{EUV Peaks from 3D Magnetic Reconnection}
\begin{document}
\begin{CJK*}{UTF8}{gbsn}

\title{Multiple Extreme Ultraviolet Peaks Attributed to Three-dimensional Magnetic Reconnection in a Long-duration Solar Flare}

\correspondingauthor{Jinhan Guo and Yu Dai}
\email{jinhan.guo@nju.edu.cn; ydai@nju.edu.cn}

\author[0009-0004-4898-638X]{Shihan Li}
\affiliation{School of Astronomy and Space Science, Nanjing University, Nanjing 210023, People's Republic of China}

\author[0000-0002-4205-5566]{Jinhan Guo}
\affiliation{School of Astronomy and Space Science, Nanjing University, Nanjing 210023, People's Republic of China}
\affiliation{State Key Laboratory of Solar Activity and Space Weather, National Space Science Center, Chinese Academy of Sciences, Beijing, People's Republic of China}

\author[0000-0002-9865-5245]{Wensi Wang}
\affiliation{CAS Key Laboratory of Geospace Environment, Department of Geophysics and Planetary Sciences, University of Science and Technology of China, 230026, Hefei, People's Republic of China}

\author[0000-0001-9856-2770]{Yu Dai}
\affiliation{School of Astronomy and Space Science, Nanjing University, Nanjing 210023, People's Republic of China}
\affiliation{Key Laboratory of Modern Astronomy and Astrophysics (Nanjing University), Ministry of Education, Nanjing 210023, People's Republic of China}

\author[0000-0003-3364-9183]{Brigitte Schmieder}
\affiliation{Centre for mathematical Plasma Astrophysics, Department of Mathematics, KU Leuven, Celestijnenlaan 200B, B-3001 Leuven, Belgium}
\affiliation{LESIA, Observatoire de Paris, CNRS, UPMC, Universit\'{e} Paris Diderot, 5 place Jules Janssen, F-92190 Meudon, France}

\author[0000-0003-1308-7427]{Jaroslav Dud\'{i}k}
\affiliation{Astronomical Institute, Czech Academy of Sciences, Fri\v{c}ova 298, 25165 Ond\v{r}ejov, Czech Republic}

\author[0000-0002-9293-8439]{Yang Guo}
\affiliation{School of Astronomy and Space Science, Nanjing University, Nanjing 210023, People's Republic of China}
\affiliation{Key Laboratory of Modern Astronomy and Astrophysics (Nanjing University), Ministry of Education, Nanjing 210023, People's Republic of China}

\author[0000-0002-4978-4972]{Mingde Ding}
\affiliation{School of Astronomy and Space Science, Nanjing University, Nanjing 210023, People's Republic of China}
\affiliation{Key Laboratory of Modern Astronomy and Astrophysics (Nanjing University), Ministry of Education, Nanjing 210023, People's Republic of China}

\begin{abstract}
Solar flares are a major driver of hazardous space weather, whose intense electromagnetic emissions and energetic particles can significantly disturb the near-Earth space environment. Therefore, understanding the physical processes during a solar flare and predicting its radiation profiles are of great importance. In this study, we analyze and model an M1.4 two-ribbon long-duration flare, whose multiple extreme-ultraviolet (EUV) emission peaks are found to correspond to different three-dimensional (3D) magnetic reconnections driven by the continuous evolution of a flux rope. In particular, the second and third peaks in the 335 {\AA} EUV channel originate from longer and higher flare loops with extended cooling times, formed by reconnection between flux-rope field lines and ambient sheared-arcade field lines ($ar\text{--}rf$) and between flux-rope field lines themselves ($rr\text{--}rf$). These results are supported by the drifting of the flux-rope footpoint (and flare ribbon) and the decrease in toroidal flux of the flux rope, as well as by the connectivity transfer of representative field lines in the magnetohydrodynamic (MHD) simulation. This work points out, for the first time, new manifestations of the 3D flare scenario in EUV light curves. On the one hand, it provides an explanation for two-ribbon late-phase flares. On the other hand, the conclusions presented here help bridge the gap between imaging observations, EUV light-curve diagnostics, and the magnetic structures of the associated coronal mass ejections.

\end{abstract}


\section{Introduction} \label{sec:intro}

Solar flares, among the most powerful energy release phenomena in the solar system, are central to our understanding of stellar magnetic activity \citep[e.g.,][]{Priest2002}. The radiation enhancement they cause can affect the Earth's ionosphere and consequently disrupt communication systems. It is well accepted that solar flares occur when magnetic energy stored in the corona is rapidly converted into plasma heating, particle acceleration, and mass motion via the process of magnetic reconnection \citep{Giovanelli1946, Sweet1958, Parker1957, Petschek1964}. 

In the framework of the standard solar flare model, also known as the CSHKP model \citep{Carmichael1964, Sturrock1966, Hirayama1974, Kopp1976}, a coronal flux rope becomes unstable and rises, forming a current sheet underneath, where magnetic reconnection occurs. This reconnection not only powers the flare emission, but also promotes the restructuring of the magnetic field, leading to the formation of new flare loops and the acceleration of coronal mass ejections \citep[CMEs;][]{Sheeley1975, Webb1987, Chen2011}. 
 
The above model is primarily described in two-dimensional (2D) context. However, in a three-dimensional (3D) scenario, magnetic reconnection differs fundamentally from the 2D case, occurring at magnetic null points \citep{Longcope2009, Pontin2016}, separators \citep{Lau1990, Priest1996, Longcope1996, Parnell2010}, and quasi-separators \citep{Titov2002}. This enables a continuous reconfiguration of the magnetic field and facilitates efficient energy release, which is crucial for understanding solar flares and coronal heating \citep{Li2021}. For example, spine-fan reconnection occurs when the null point collapses into a current sheet. Alternatively, torsional motions can drive reconnection along the spine or the fan of the null point \citep{Priest2009, Prasad2018, Nayak2019}. Besides nulls, other important sites for reconnection include separators and quasi-separatrix layers (QSLs). Separators are formed by the intersection of two separatrix surfaces, where magnetic field line mapping becomes singular, and they have been identified in certain flare events \citep[e.g.,][]{Titov2012}. Meanwhile, the spatial correspondence between flare ribbons and QSLs has been widely observed, indicating that QSLs are also favorable locations for magnetic reconnection to occur \citep{Priest1995, Demoulin1996A&A, Demoulin2006, Liu2016}. For more details about 3D magnetic reconnection, please refer to the recent review of \citet{Dudik2025}.

Based on the 3D magnetic reconnection pattern, a 3D flare--CME model has been proposed \citep{Aulanier2012, Janvier2015, Aulanier2019}. Within this 3D model, two new distinct reconnection geometries have been identified. The first involves reconnection between inclined sheared-arcade field lines ($a$, arcade) and the rising flux-rope field lines ($r$, flux rope), producing new flux-rope field lines eq89¢ib and flare-loop field lines ($f$, flare loop); this process is referred to as $ar\text{--}rf$ reconnection. Such reconnection can lead to the drifting of flux-rope footpoints, the formation of non-coherent magnetic structures without a common CME axis \citep{Aulanier2019, Dudik2019, Guojh2024, Guojh2025}, and the development of saddle-shaped flare loops \citep{Lorincik2021}. The second geometry occurs between the flux-rope field lines themselves and is termed $rr\text{--}rf$ reconnection. This process transfers the toroidal flux to poloidal flux, and results in the formation of a more twisted flux rope \citep{Aulanier2019, Xing2020}. Notably, MHD simulations \citep{Aulanier2019} depict a temporal overlap and transition between these reconnection geometries as the reconnection-driving flux rope rises and continuously evolves.

In observations, flares are often classified by their soft X-ray (SXR) light curves observed by missions like the Geostationary Operational Environmental Satellite (GOES), which exhibit an impulsive rise and a gradual decay \citep[e.g., ][]{Woods2011}. Long-duration events (LDEs), in particular, are characterized by their slow decay phases, often lasting for hours \citep{Kahler1977}. These events are typically associated with large-scale arcades of post-flare loops and are almost invariably accompanied by fast CMEs. The extended energy release in LDEs has been traditionally attributed to the slow cooling of plasma heated during the impulsive phase \citep[e.g.,][]{Cargill1995}. However, evidence from H$\alpha$ and subsequent C~\textsc{iv} and SXR data observations revealed gentle chromospheric evaporation with sustained upflows \citep{Schmieder1987, Schmieder1990}, supporting an energy supply process during the gradual phase. Moreover, multi-peaked EUV light curves indicate a complex, multi-stage energy release rather than a single impulsive injection \citep[e.g.,][]{Liu2015, Li2018}.

It is important to note that the observed delays between the multiple emission peaks may not exclusively reflect intermittencies in the magnetic reconnection process itself. Instead, the temporal profile of a flare can be modulated by thermodynamic processes, particularly the different cooling timescales of flare loops formed by different reconnection mechanisms. For instance, \citet{Lorincik2021} suggested that $ar\text{--}rf$ reconnection involving the flux rope can produce flare loops that are longer, higher, and more inclined than those formed by $aa\text{--}rf$ reconnection. Since the cooling time scales with loop length, the different length between the compact loops formed early (e.g., via $aa\text{--}rf$) and the extended loops formed later (e.g., via $ar\text{--}rf$) can introduce substantial time lags in EUV emissions. Therefore, combining the 3D magnetic reconnection scenario with the thermodynamic response is crucial for a more comprehensive understanding of the evolution of LDEs. To this end, we analyze an M1.4 long-duration flare on 2011 August 2. Combining observations from the Solar Dynamics Observatory \citep[SDO;][]{Pesnell2012} with data-constrained magnetohydrodynamic (MHD) simulations, we identify multiple peaks in the EUV light curves, and associate them with different 3D reconnection mechanisms, which are driven by a continuously evolving flux rope system. This combined approach helps bridge the gap between observed EUV emission signatures and underlying 3D magnetic evolution.

\section{Observations}\label{sec:obs}

\subsection{Light Curves of the Solar Flare}

In this study, we focus on an M1.4-class solar flare that occurred on 2011 August 2 in NOAA Active Region (AR) 11261. For this event, \citet{Wang2025} have investigated the evolution of a coronal flux rope carrying a substantial non-neutralized current, from its formation through eruption.

\begin{figure}[ht!]
\centering
\includegraphics[scale=0.9]{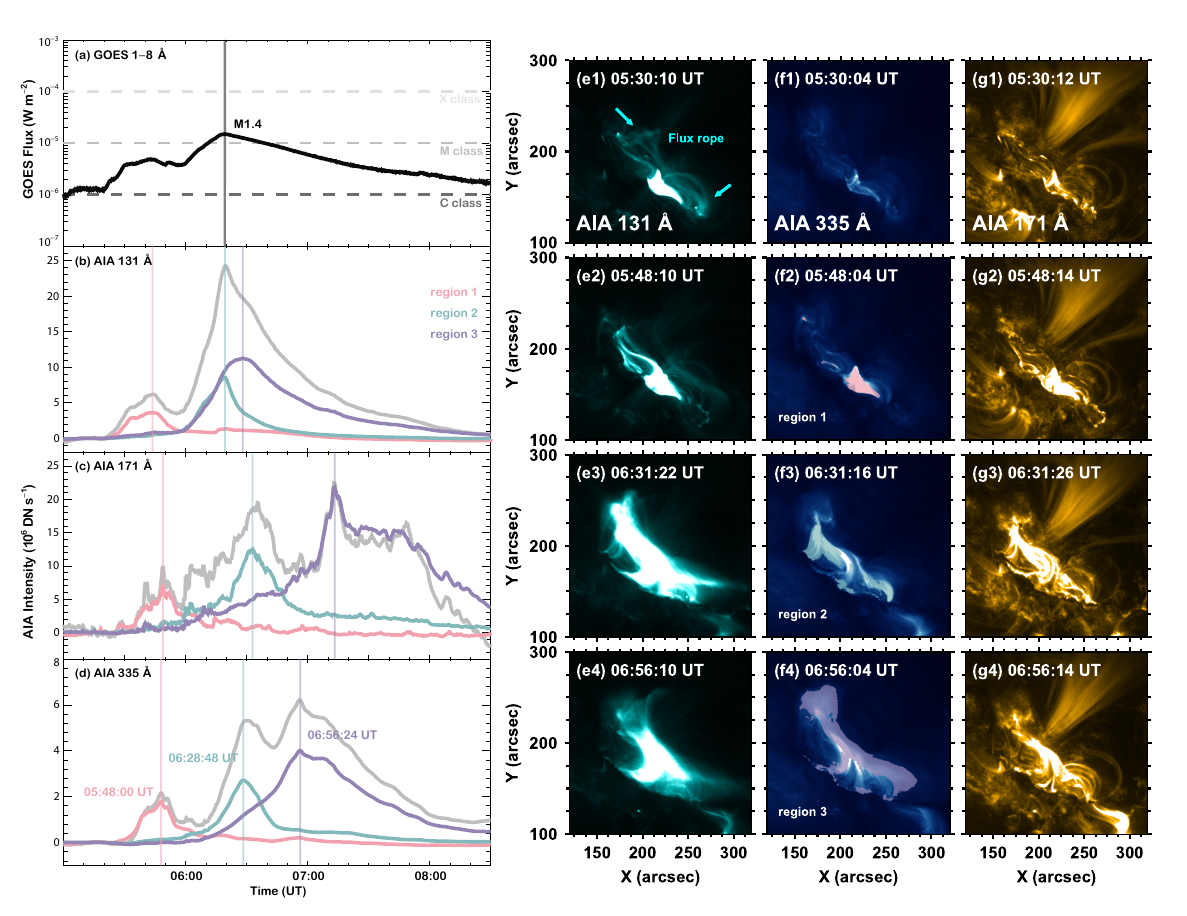}
\caption{General evolution of the flare. Left: Time profiles of the GOES 1--8 {\AA} flux (a), and light curves from AIA 131 {\AA} (b), AIA 171 {\AA} (c) and AIA 335 {\AA} (d) integrated over the entire active region (gray) as well as over Region 1, Region 2, and Region 3 (colored). The background is subtracted from each light curve using the average intensity during a quiescent period. Vertical dashed lines indicate prominent peaks in the GOES flux and AIA intensities. Right: snapshots of the flare evolution in AIA 131 {\AA} (e1)--(e4), 335 {\AA} (f1)--(f4), and 171 {\AA} (g1)--(g4). The colored areas in panels (f2), (f3) and (f4) mark the three regions of interest. Several characteristic features are also highlighted; see the text for details. Note that the AIA subregion covers a field of view of $200^{\prime\prime} \times 200^{\prime\prime}$, enclosing the flare-hosting active region.}
\label{figure1}
\end{figure}

Figure \ref{figure1}(a) shows the time profile of GOES 1--8 {\AA} SXR flux for this event. The long-duration flare begins at approximately 05:19 UT and peaks at 06:19 UT. The GOES light curve exhibits two conspicuous peaks, the first of which can be identified as pre-flare brightenings. Observations taken with the Atmospheric Imaging Assembly \citep[AIA;][]{Lemen2012} provide full-disk images of the transition region (TR) and corona in ten passbands with a pixel scale of 0.6{\arcsec} and a cadence of 12 or 24 s. Figures \ref{figure1}(e)--(g) display the evolution of the flare in AIA 131 {\AA}, 335 {\AA}, and 171 {\AA}. Meanwhile, Figures \ref{figure1}(b)--(d) present the light curves of the entire active region in these three passbands, which are obtained by integrating the intensities over all pixels within the rectangular area defined by $x \in [120\arcsec, 320\arcsec]$ and $y \in [100\arcsec, 300\arcsec]$. The resulting profiles reveal three distinct peaks in both the AIA 171 {\AA} and 335 {\AA} channels, which may correspond to multiple magnetic reconnection events.

According to previous studies, a flux rope (marked in Figure \ref{figure1}(e1)) formed from a coherent, low-lying bright structure and was driven by slow photospheric evolution \citep{Wang2025}. At different temperatures sampled by the AIA passbands, pre-flare loops can be identified in Figures \ref{figure1}(e1)--(g1). As the flux rope erupts and rises, the pre-flare loop system exhibits enhanced brightness (see Figures \ref{figure1}(e2)--(g2)). As the arcade continues to develop, a saddle-like structure with two cantles becomes visible (Figure \ref{figure1}(e3)). This configuration features higher-lying flare loops at its extremities and lower ones in the center, resembling the cantles of a saddle in horseback riding terminology \citep[see][]{Lorincik2021}. In the final row, the saddle-shaped loops gradually fade and are eventually replaced by loops with different topological structures. This sequence of images reveals that the post-flare loops appear as spatially discrete structures with varying brightness.

We examine whether the three sets of flare loops observed in the images correspond to the three peaks in the AIA 171 {\AA} and 335 {\AA} light curves. Here, we improve the method of \citet{Chen2020} to estimate the emission areas of different regions as follows. First, two AIA 335 {\AA} images taken at the first and second peaks are selected, and the latter is subtracted from the former. In the resulting difference image, the Region 1 (Region 2) corresponds to pixels above (below) the mean of all positive (negative) pixels. Similarly, by subtracting images between the first and third peaks, another set of Region 1 and Region 3 is obtained. The same procedure is applied between the second and third peaks to derive another Region 2 and Region 3. Finally, the common overlapping areas of the two respective versions of each region are taken as the final Regions 1, 2, and 3, and a median filter is applied to reduce noise. The identified regions are indicated by the pink area (Region 1) in Figure \ref{figure1}(f2), the green area (Region 2) in Figure \ref{figure1}(f3), and the purple area (Region 3) in Figure \ref{figure1}(f4), respectively. To evaluate the robustness of the region extraction, we computed the AIA intensity profiles separately within these three subregions. As shown in Figures \ref{figure1}(b)--(d), the emissions from the extracted regions correspond well to the three distinct peaks in the light curves. Moreover, within each region, the emission peak occurs earlier in hotter spectral lines and later in cooler ones, possibly reflecting a cooling process.

It is noted that Region 1 (pink area), which corresponds to the first peak, covers a relatively small area located close to the core of the active region. After the onset of the impulsive phase of the flare, this region was likely overwhelmed by subsequent brighter and more widespread emissions, making it difficult to accurately analyze its independent evolution. More importantly, the pre-flare activity associated with Region 1 is not the focus of our study, as it has already been thoroughly investigated by \citet{Wang2025}. Therefore, we concentrate on explaining the physical processes responsible for the second and third peaks.

\subsection{Identifying 3D Magnetic Reconnection Geometries}\label{sub2_2}

\begin{figure}[ht!]
\centering
\includegraphics[scale=1.1]{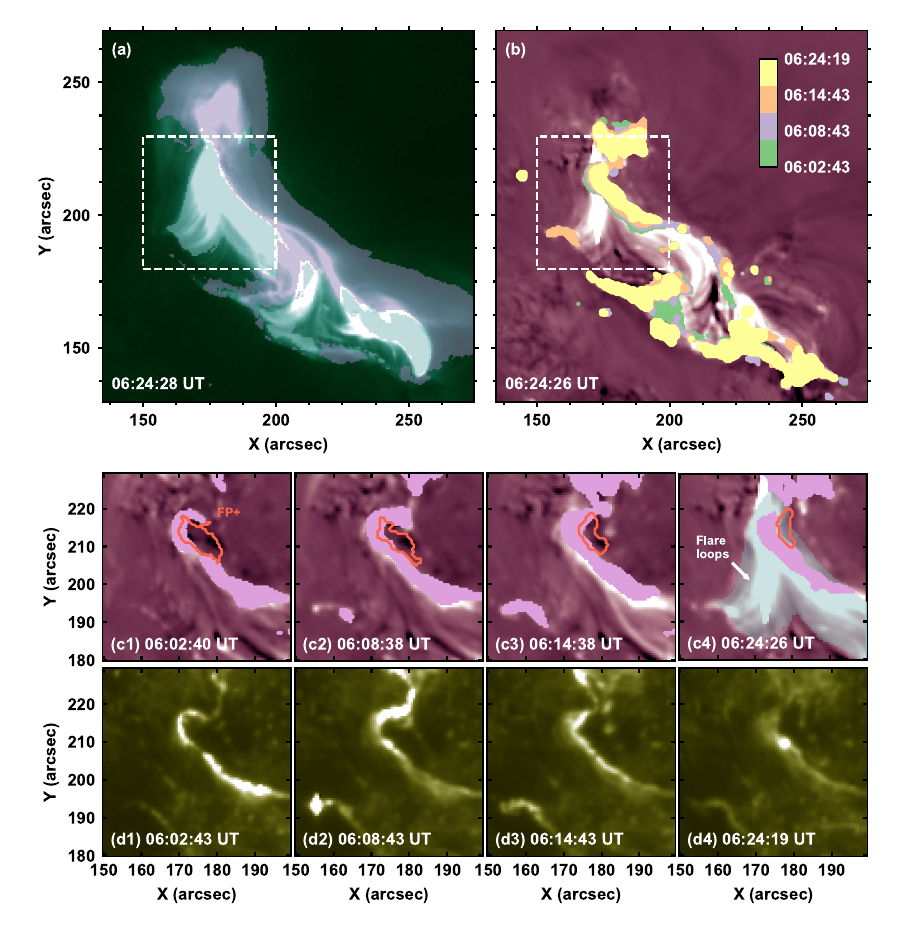}
\caption{Flare ribbon observations in AIA filter channels. Panel (a) presents a zoomed-in view of the flare-hosting active region in AIA 94 {\AA}, with Region 2 and Region 3 overlaid as in Figure \ref{figure1}. Panel (b) displays the traced flare ribbons at 06:02, 06:08, 06:14, and 06:24 UT. Colored dots denote the locations shown in panels (c1)--(c4) (purple dots). Selected characteristic structures are highlighted; see the text for further details. A zoomed-in view of the area within the dashed boxes in panels (a) and (b) is shown in panels (c) and (d). Panels (d1)--(d4) illustrate the temporal evolution of the flare ribbons in AIA 1600 {\AA}, corresponding to the times shown in panels (c1)--(c4). Red contours in panels (c) outline the positive footpoint, labeled as ``FP+". The animation illustrates the drifting of the flare ribbon and the associated formation of flare loops near the footpoint across its 6-second duration.
\\
(An animation of this figure is available.)}
\label{figure2}
\end{figure}

The evolution of flare ribbons in the chromosphere helps understand the process of magnetic reconnection in the corona. Based on \citet{Wang2017}, we identify flare ribbons in the AIA 1600 {\AA} passband by selecting all brightening pixels within the active region that satisfy the following criteria: intensity values are 4--6 times greater than the pre-flare average brightness, and the brightening sustains for several minutes. The ribbons detected at four different times are overplotted in Figure \ref{figure2}(b) as dot symbols in different colors. Within the region enclosed by the dashed rectangle, the flare ribbons exhibit a westward shift over time.

To further analyze the flare ribbon evolution, we examine this zoomed-in region using AIA 211 {\AA} differential images at four selected times. The first differential image (Figure \ref{figure2}(c1)) reveals a dimming region (for the identification of dimming regions, see Appendix \ref{appA}) encircled by hook-shaped flare ribbons (marked by purple dots). Models suggest that flux-rope footpoints are surrounded by J-shaped current ribbons \citep{Demoulin1996, Pariat2012}, produced by the intersection of QSLs with the photosphere \citep{Aulanier2012}. Such structures correspond to UV flare ribbons \citep{Dudik2014, Janvier2016, Zhao2016, Lorincik2019} and are sometimes linked to coronal dimmings \citep{Dissauer2018}.

At three subsequent time points (Figure \ref{figure2}(c)), with the corresponding AIA 1600 {\AA} ribbon evolution shown in Figure \ref{figure2}(d), first, the hook-shaped flare ribbon and flux-rope footpoint (we name it FP+ because of its location in positive polarity according to Appendix \ref{appA}) both exhibit drifting. Second, propagating hook-shaped ribbon sweeps across the initial flux rope footpoint location (Figure \ref{figure2}(c1)), leading to the formation of new flare loops rooted near FP+. By approximately 06:24 UT (Figure \ref{figure2}(c4)), clear flare loops are visible, consistent with $ar\text{--}rf$ reconnection \citep{Aulanier2019}. Furthermore, FP+ exhibits a contraction process, indicating the occurrence of internal reconnection within the flux rope, which may correspond to the $rr\text{--}rf$ reconnection \citep{Xing2020}.

Since three distinct regions were previously identified, we overplot Region 2 and Region 3 on an AIA 94 {\AA} image to locate where the $ar\text{--}rf$ reconnection occurs (Figure \ref{figure2}(a)). The zoomed-in view shows that both the area swept by the flare ribbons and the newly formed flare loops lie within Region 2, as emphasized in Figure \ref{figure2}(c4). This process is absent in Region 3; therefore, we suggest that the $ar\text{--}rf$ reconnection contributes to the second peak in the light curve.

\begin{figure}[ht!]
\centering
\includegraphics[scale=0.9]{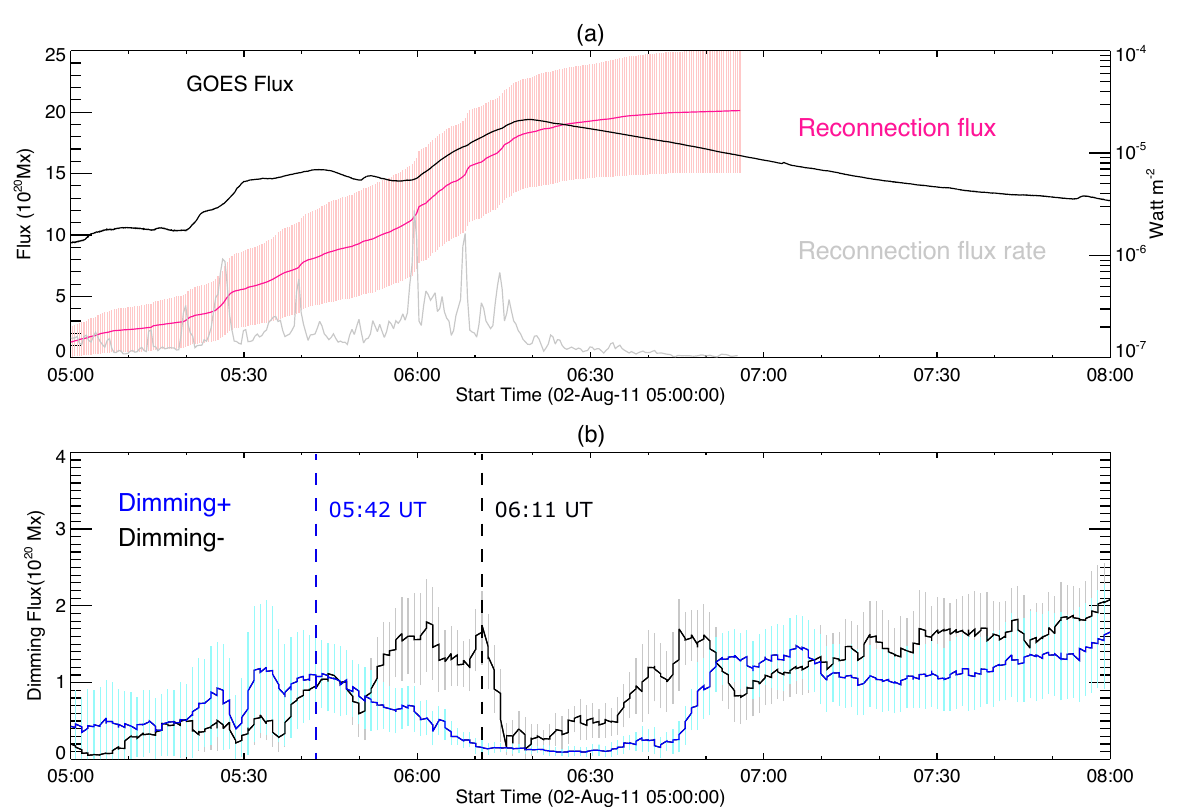}
\caption{Evolution of the reconnection flux and flux-rope footpoints. Panel (a) presents the soft X-ray light curve (black), the magnetic reconnection flux (pink; shaded area indicating uncertainties), and the time derivative of the reconnection flux (gray). Panel (b) shows the magnetic fluxes of the northern positive (blue) and southern negative (black) dimming regions. The blue and black dashed lines mark the onset of flux decrease in the northern positive and southern negative dimming regions, respectively.}
\label{figure3}
\end{figure}

To further investigate the evolution of the flux rope during the eruption, following \citet{Wang2017, Wang2025}, we compute the reconnection flux and its rate (Figure~\ref{figure3}(a)) based on the evolution of flare ribbons, and the magnetic flux of the flux-rope footpoints (Figure~\ref{figure3}(b)) using conjugate dimming regions (Figure~\ref{fig_app1}). As shown in Figure~\ref{figure3}(a), reconnection flux continues to increase until 06:56 UT, with its rate peaking around 05:58 UT and dropping to zero near 06:20 UT. It is noted that the reconnection rate exhibits fluctuations, consistent with the fragmented flare ribbons observed in high-resolution data \citep{Faber2025}. Regarding the evolution of magnetic flux of flux-rope footpoints, both the northern positive and southern negative dimming regions display a pattern of rising, then falling, and rising again, with decreases occurring around 05:42 UT and 06:11 UT, respectively. In general, increases and decreases in magnetic flux are interpreted as corresponding to the $aa\text{--}rf$ and $rr\text{--}rf$ reconnection geometries in the 3D flare model \citep{Jiang2021}. The later increase in the magnetic flux of the dimming regions reflects the recovery phase following the eruption. However, attributing a specific magnetic flux change signature to the $ar\text{--}rf$ reconnection remains challenging. While $ar\text{--}rf$ reconnection is expected to alter the connectivity of the flux rope \citep{Aulanier2019}, it does not change the toroidal flux \citep{Xing2020}. Therefore, it may not produce an unambiguous imprint on the magnetic flux evolution curve, unlike the clearer signatures of $aa\text{--}rf$ (increase) and $rr\text{--}rf$ (decrease) reconnections.

Collectively, Region 1 is associated with the $aa\text{--}rf$ reconnection, evidenced by the overall increase in magnetic flux. The observed drifting and contraction of the footpoint (FP+) and the temporal evolution of the magnetic flux in the dimming regions provide evidence for the sequence of reconnection processes. Specifically, the $ar\text{--}rf$ reconnection (and potential early $rr\text{--}rf$ reconnection), and the subsequent $rr\text{--}rf$ reconnection correspond to Region 2 and Region 3, respectively. The sequential brightening of distinct regions (Regions 1--3) suggests that the evolving flux rope sequentially interacts with different magnetic topological domains during its ascent.

\subsection{Formation Mechanisms of Multiple EUV Peaks}

Here, we investigate the thermal evolution during the eruption to explore the formation mechanisms of multiple EUV peaks. Using six AIA coronal passbands, we perform differential emission measure (DEM) analysis on the active region. To improve the signal-to-noise ratio, we first spatially rebin each AIA image by a factor of 2 and temporally average every five consecutive images to form the input data. We then apply a sparse inversion algorithm \citep{Cheung2015, Su2018} over a temperature range of $\log(T/\mathrm{K}) \in[5.5, 7.6]$ with a grid spacing of 0.05 dex.

\begin{figure}[ht!]
\centering
\includegraphics[scale=0.9]{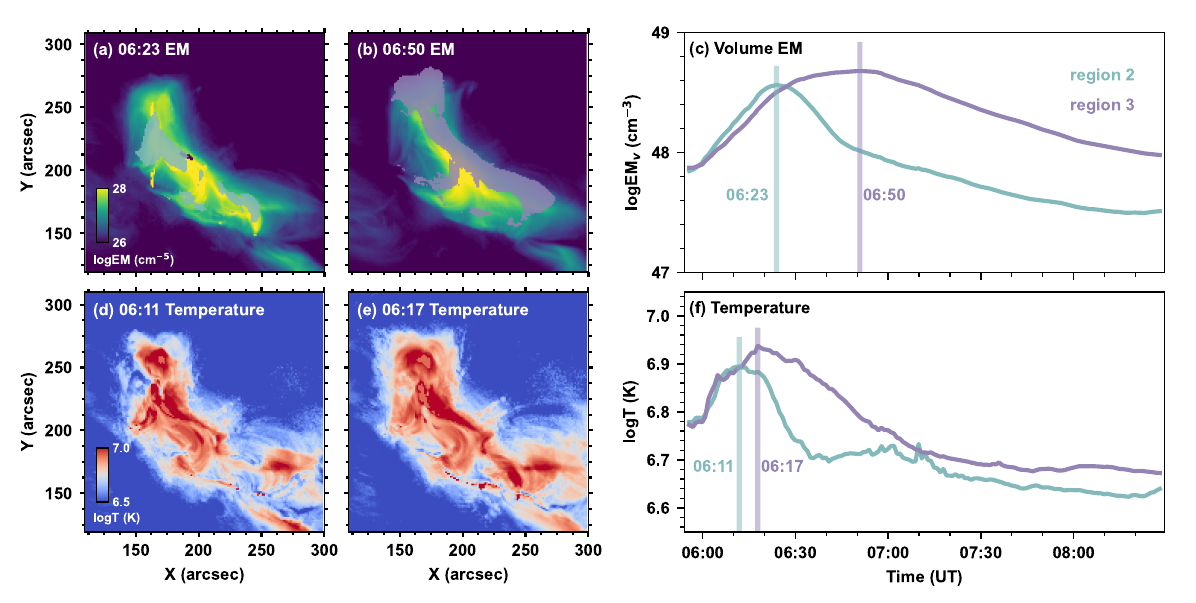}
\caption{DEM inversion results for the flare. Top: EM maps of the active region at two selected times when Region 2 and Region 3 (delineated by the colored areas in panels (a) and (b), respectively) are clearly defined. Panel (c) represents the temporal evolution of the volume emission measure (EM) for Region 2 and Region 3 (distinguished by color). Vertical lines mark the respective times of peak volume EM for each region. Bottom: DEM-weighted temperature maps of the active region at two selected times (panels (d) and (e), respectively). Panel (f) represents the temporal evolution of the DEM-weighted temperature for Region 2 and Region 3, with vertical lines marking the peak times for each region.}
\label{figure4}
\end{figure}

From the DEM results, we can obtain the volume EM ($EM_V$) and DEM-weighted temperature ($\bar{T}$) over a specific region \citep[see][]{Li2024}. Figures \ref{figure4}(a) and (b) present the EM maps at two distinct times corresponding to the peaks in the EM temporal profiles, while Figures \ref{figure4}(d) and (e) show the DEM-weighted temperature maps at the times of their respective temperature peaks. To quantitatively track the evolution of the two regions, the temporal changes in their volume EM and DEM-weighted temperature are plotted in Figures \ref{figure4}(c) and (f). The light curves of the temperature for Region 2 and Region 3 in Figure \ref{figure4}(f) exhibit similar trends during the rise phase but show a 6-minute difference in the peak time. This temporal offset may reflect a multi-stage energy release process rather than a single heating input. Furthermore, their distinct cooling behaviors after the peaks likely reflect differences in the physical properties of two regions.

On the one hand, the DEM-weighted temperatures of Region 2 and Region 3 exhibit a decreasing trend after reaching their respective peaks (Figure \ref{figure4}(f)). Although small fluctuations are observed in the Region 2 curve, they do not sustain the temperature at a high level over an extended period nor produce a new significant peak. That is, these fluctuations do not alter the overall trend of the temperature curve. Therefore, we suggest that there is no clear evidence of additional heating in either region. 
On the other hand, by approximating each region with a set of identical characteristic loops, we can theoretically estimate its overall cooling time. Here, we adopt the improved loop cooling time formula proposed by \citet{Cargill1995} and modified by \citet{Li2024}:

\begin{equation}
\tau_{\mathrm{cool}} = 2.04\times 10^{-2}L^{5/6}T_{0}^{-1/6}n_{0}^{-1/6}\ \ \mathrm{[s]},\label{tcool_eqn}
\end{equation}
where $L$ is the characteristic loop half-length; $T_0$ and $n_0$ are the temperature and density at the onset of cooling. We approximate $L=S^{1/2}$, where $S$ is the projected area of each region. $T_0$ is set to the peak DEM-weighted temperature of each region, and $n_0$ is derived from the corresponding volume emission measure as $(EM_V/V)^{1/2}$, with the volume given by $V=S^{3/2}$. 

Based on this approach, the theoretically estimated cooling times for Region 1, Region 2, and Region 3 are about 25 minutes, 56 minutes, and 76 minutes, respectively. Such differences can be expected from the increasing loop length across the three regions, which serves as a primary factor in determining the loop cooling time. Furthermore,  despite a rather continuous transition of the energy release between the considered regions (as inferred form Figure \ref{figure4}(f)), their differences in cooling time are sufficiently large to account for the observed time delays between multiple EUV peaks (Figures \ref{figure1}(b)--(d)). In passing, we note that the above approach provides a conservative estimate for the actual loop lengths, which may be substantially longer. Consequently, the difference in cooling times could be even more significant than estimated here.

\section{Data-constrained MHD Simulation}\label{sec:num}

\subsection{Modeling Setup}

To investigate the reconnection process during the eruption, we adopt a data-constrained MHD simulation to study the evolution of magnetic topology based on zero-$\beta$ assumption, which neglects the effects of gravity, gas pressure, and energy equation. Following \citet{Guo2019a}, the governing equations are as follows:

\begin{eqnarray}
 && \frac{\partial \rho}{\partial t} +\nabla \cdot(\rho \boldsymbol{v})=0,\label{eq1}\\
 && \frac{\partial (\rho \boldsymbol{v})}{\partial t}+\nabla \cdot(\rho \boldsymbol{vv}-\boldsymbol{BB})+\nabla (\frac{\boldsymbol{B}^2}{2})=0,\label{eq2}\\
 && \frac{\partial \boldsymbol{B}}{\partial t} + \nabla \cdot(\boldsymbol{vB-Bv})=0,\label{eq3}
\end{eqnarray}
where ${\rho}$, $\boldsymbol{v}$ and $\boldsymbol{B}$ represent the density, velocity and magnetic fields, respectively. 

The initial magnetic fields are obtained from coronal magnetic-field extrapolation using the pipeline of \citet{Guo2019b}, which consists of (1) preprocessing of the vector magnetogram and (2) numerical reconstruction of the 3D coronal magnetic fields. In the preprocessing step, the bottom vector magnetic field uses observations at 05:24 UT on 2011 August 2, provided by the Helioseismic and Magnetic Imager \citep[HMI;][]{Scherrer2012, Schou2012}. The 180 degree ambiguity caused by the Zeeman effect is resolved, and projection effects are corrected by remapping to the local Cartesian coordinate system, as described in \citet{Guo2016a}. Subsequently, the coronal magnetic fields are extrapolated from the potential-field model using the $B_{z}$ component with the Green's function. A twisted flux rope is then inserted based on the regularized Biot-Savart laws (RBSL) method \citep{Titov2018}. The flux-rope path is constrained by the observed hot channel, while parameters such as the minor radius and toroidal magnetic flux are estimated from its footpoints \citep{Wang2025}.

The initial plasma density is prescribed by a stratified hydrostatic solution (d$(\rho T)$/d$h$ = $-\rho g$), with the temperature distribution simplified as a stepwise function to represent the cold, dense chromosphere and the hot, tenuous corona. For the bottom boundary conditions, the 3D magnetic fields are constrained by the observed vector magnetograms, while the density is fixed to its initial value. On the other boundaries, a zero-gradient extrapolation is applied to implement open boundary conditions.

The above partial differential equations are numerically solved with the open-source Message Passing Interface Adaptive Mesh Refinement Versatile Advection Code \citep[MPI-AMRVAC\footnote{http://amrvac.org},][]{Xia2018, Keppens2023}. The computational domain is ${[x_{min},x_{max}]\times[y_{min},y_{max}]\times[z_{min},z_{max}]=[-177,177]\times[-147,147]\times[1,354]}$ Mm, which is resolved by  ${240\times200\times240}$ cells, with a uniform gridding. We use an HLL Riemann solver for flux reconstruction, a three-step Runge-Kutta time stepper, and a third-order Cada limiter.                                                                            
\subsection{Numerical Results}
\subsubsection{Global Evolution and Comparison with Observations}
\begin{figure}[ht!]
\centering
\includegraphics[scale=0.5]{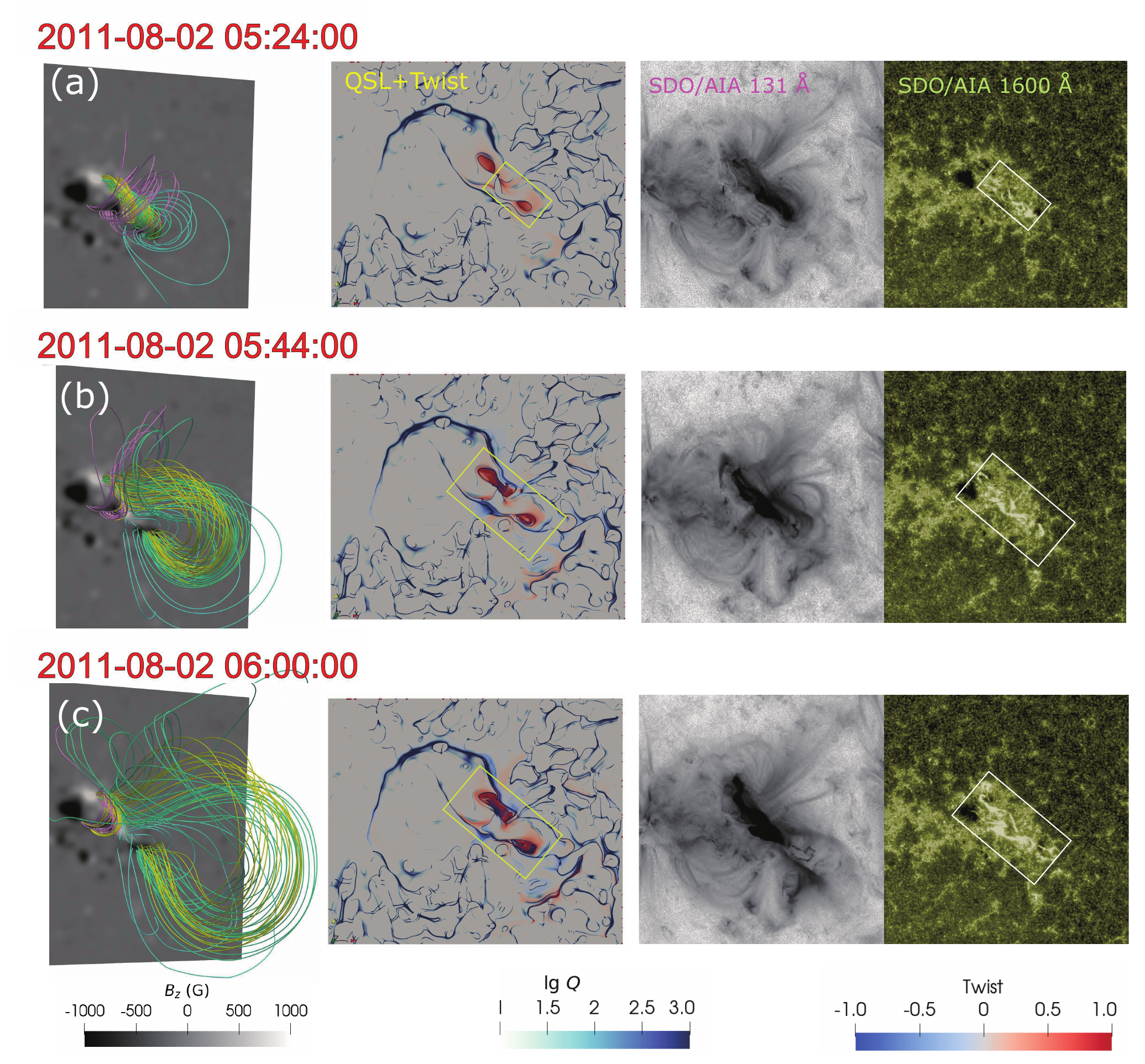}
\caption{Top-view snapshots of the data-constrained numerical simulation and their comparison with AIA imaging observations for the 2011 August 2 flare. The three rows from the top to bottom display the observations and simulations at 05:24, 05:44, and 06:00 UT. The first column shows the top views of the 3D magnetic field configurations revealed from the simulation, where the pink and cyan field lines correspond to overlaying arcades at the eruption onset, and the yellow and green field lines show the flux rope with different topological connectivity. The second column presents the QSLs and twist distributions on the bottom plane. The third and fourth column present AIA 131 {\AA} and 1600 {\AA} images, respectively. The yellow (second column) and white (fourth column) boxes in QSLs and 1600 \AA\ images outline the regions of flare ribbons. An animation showing the evolution of the simulation from 05:24 to 06:14 UT is available. The real-time duration of the animation is 5.3 s.
\\
(An animation of this figure is available.)}
\label{figure5}
\end{figure}

Figure \ref{figure5} presents a side-by-side comparison of data-constrained MHD simulation results (left two columns) and simultaneous SDO/AIA observations (right two columns) at three key time points during the 2011 August 2 M1.4 flare. The analysis focuses on the evolution of the magnetic topology and its correspondence with observed flare features, supporting a 3D magnetic reconnection scenario.

In the simulation, the initial magnetic configuration consists of overlying arcade fields depicted in pink and cyan, and flux rope fields shown in yellow and green. We also use the method proposed by \citet{Zhang2022} to calculate the squashing factor Q \citep{Priest1995, Demoulin1996, Titov2002} and twist number, which help characterize flux rope boundaries and locate regions favorable for magnetic reconnection \citep{Guo2017, Guo2021a, Guo2023}. On the bottom boundary, the Q-map exhibits a pair of hook-shaped features, which surround regions of high twist that mark the footpoints of the flux rope. Observations in AIA 131 {\AA} show the initial brightening of hot flare loops, while the AIA 1600 {\AA} image captures the early development of two primary flare ribbons (Figure \ref{figure5}(a)). The initial magnetic fields in simulations are comparable to observations. On the one hand, the simulation shows that the modeled flux rope closely resembles the observed hot channel. On the other hand, the pre-flare ribbons (white rectangle) are consistent with the straight part of two-ribbon QSLs in the simulation (yellow rectangle).

By the next time point, the simulation shows significant evolution. The yellow and green field lines, representing the flux rope, have clearly risen and expanded (Figure \ref{figure5}(b)). During this period, part of the pink arcades and the yellow flux rope transform into flare loops connecting the northern hook-shaped region (footpoint of the flux rope) and the southern footpoint of the original arcade. The Q-map becomes more pronounced and extended, with the hook-shaped structure evolving and the region of high twist number expanding, indicating actively progressing magnetic reconnection. The arcade field lines reconnect with the flux rope field lines near the footpoint, resulting in footpoint drifting. The northern footpoint exhibits a more pronounced drifting, as evidenced by the more significant displacement of the hook-shaped structure outlined by the high Q and twist regions on the northern side, corresponding to the $ar\text{--}rf$ reconnection process. This corresponds to the observed drifting of the northern dimming regions, surrounded by hooked flare ribbons, as shown in Figure~\ref{figure2}. Moreover, the AIA 131~{\AA} image displays a well-defined set of bright hot loops, whereas the AIA 1600~{\AA} image shows further development and separation of the flare ribbons. Similar separation motions are also seen in the QSLs between the two high-twist footpoints in the simulation. A detailed comparison at this moment through image superposition is presented in Appendix \ref{appB}.

In the final stage (Figure \ref{figure5}(c)), the flux rope rises higher and becomes more twisted. An indicator of the $rr\text{--}rf$ reconnection responsible for this twist enhancement is the transformation of part of the green flux-rope field lines into flare loops with different connectivity; their footpoints are located between the two hook-shaped regions. Meanwhile, the original overlying pink arcades have largely formed into flare loops. The photospheric QSLs have expanded, especially at the edge of the hook-shaped structure, and both the squashing factor and magnetic twist have increased significantly, reflecting ongoing reconnection near the flux-rope footpoints. Observationally, the AIA 131 {\AA} emission shows a more mature, large-scale post-flare loop arcade compared to the previous time, and the AIA 1600 {\AA} image reveals fully separated, brighter flare ribbons.

In summary, the side-by-side comparison across these three time points demonstrates consistent agreement between the data-constrained simulation and observations. The model successfully reproduces (1) the expansion and rise of a flux rope (the morphology and eruption direction of hot channel);  (2) the evolution of hook-shaped QSLs that spatially match the observed flare ribbons (separation motions), and (3) the sequential formation of hot post-flare loops. This correspondence validates that the long-duration energy release in this event was driven by a continuous evolution of 3D magnetic reconnection processes, involving sustained interactions between an erupting flux rope and the overlying arcade, and between the flux-rope field lines themselves.

\begin{figure}[ht!]
\centering
\includegraphics[scale=0.7]{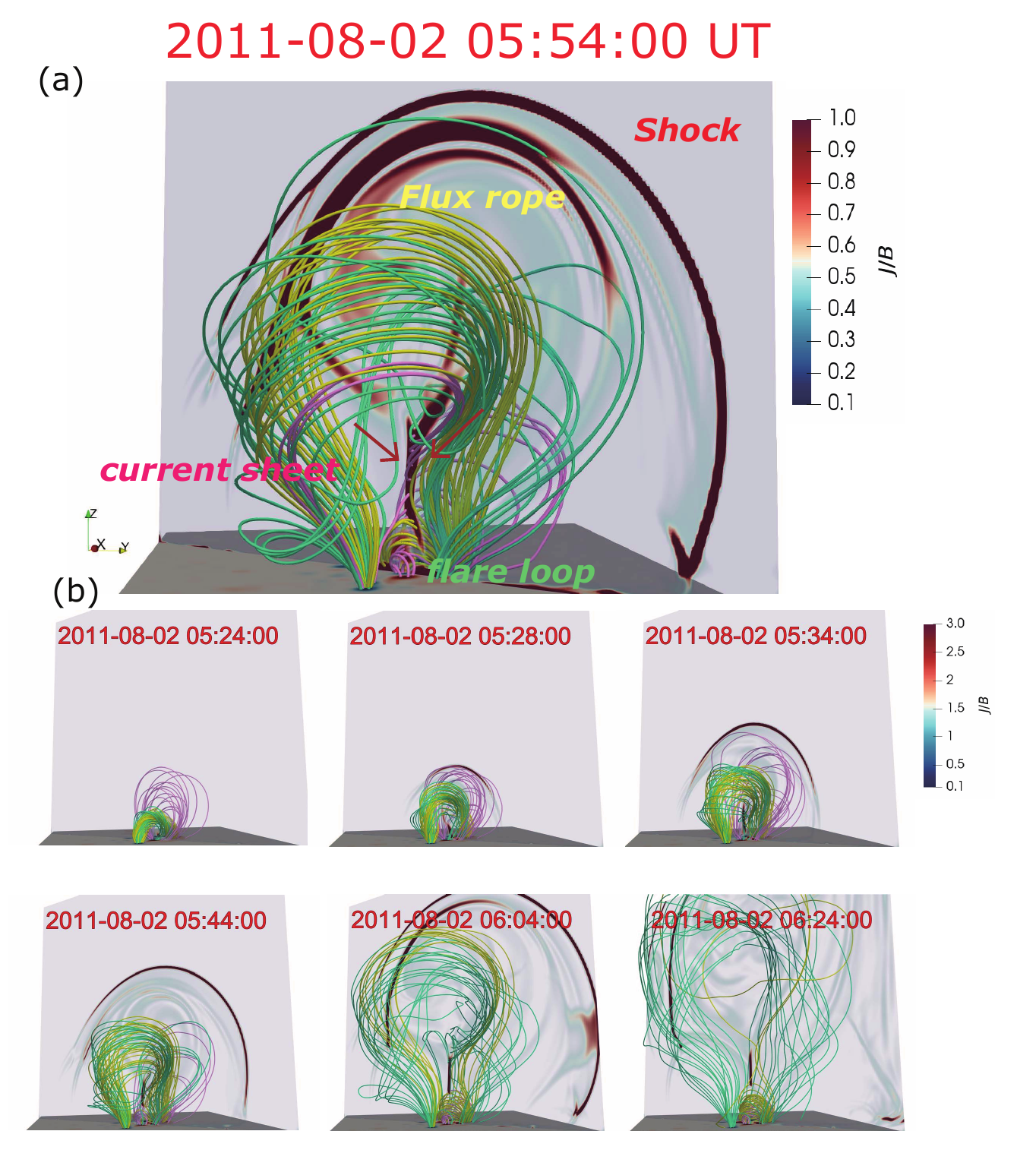}
\caption{Similar to Figure \ref{figure5} but for the side view, with the magnetic field lines along with the distribution of J/B. An animation showing the side-view evolution of the simulation from 05:24 to 06:24 UT is available. The real-time duration of the animation is 6.2 s.
\\
(An animation of this figure is available.)}
\label{figure6}
\end{figure}

From the side perspective, the simulation reveals additional details. The scalar metric J/B, which highlights abrupt changes in magnetic connectivity, is commonly used to identify current sheets where magnetic reconnection is likely to occur \citep{Gibson2006, Fan2007, Jiang2016}. Figure \ref{figure6}(a) displays a snapshot of the simulation at 05:54 UT, where a thin, sheet-like structure is visible between the flux rope and the pink arcades, surrounded by an elongated region of elevated J/B values. This feature corresponds to a current sheet that facilitates magnetic reconnection. Within this region, the flux rope and the nearby pink arcades undergo $ar\text{--}rf$ typed reconnection, leading to the formation of flare loops below the current sheet.

Figure \ref{figure6}(b) tracks the eruption dynamics from 05:24 UT to 06:24 UT. As the flux rope expands outward, it progressively stretches the underlying magnetic fields, resulting in the elongation and intensification of the current sheet, particularly evident by 06:04 UT. The field lines exhibit increased dynamical behavior, with the yellow and green flux ropes interacting closely. Throughout this period, the configuration evolves such that conditions become increasingly favorable for the onset of reconnection between the flux ropes ($rr\text{--}rf$ reconnection), which could contribute to the formation of a more twisted flux rope and post-flare loops beneath it. Thus, the evolution captured in the simulation illustrates the transition from a dominant $ar\text{--}rf$ reconnection topology to $rr\text{--}rf$ reconnection topology as the eruption progresses.

\subsubsection{3D Magnetic Reconnection in the Eruption Process}

\begin{figure}[ht!]
\centering
\includegraphics[scale=0.8]{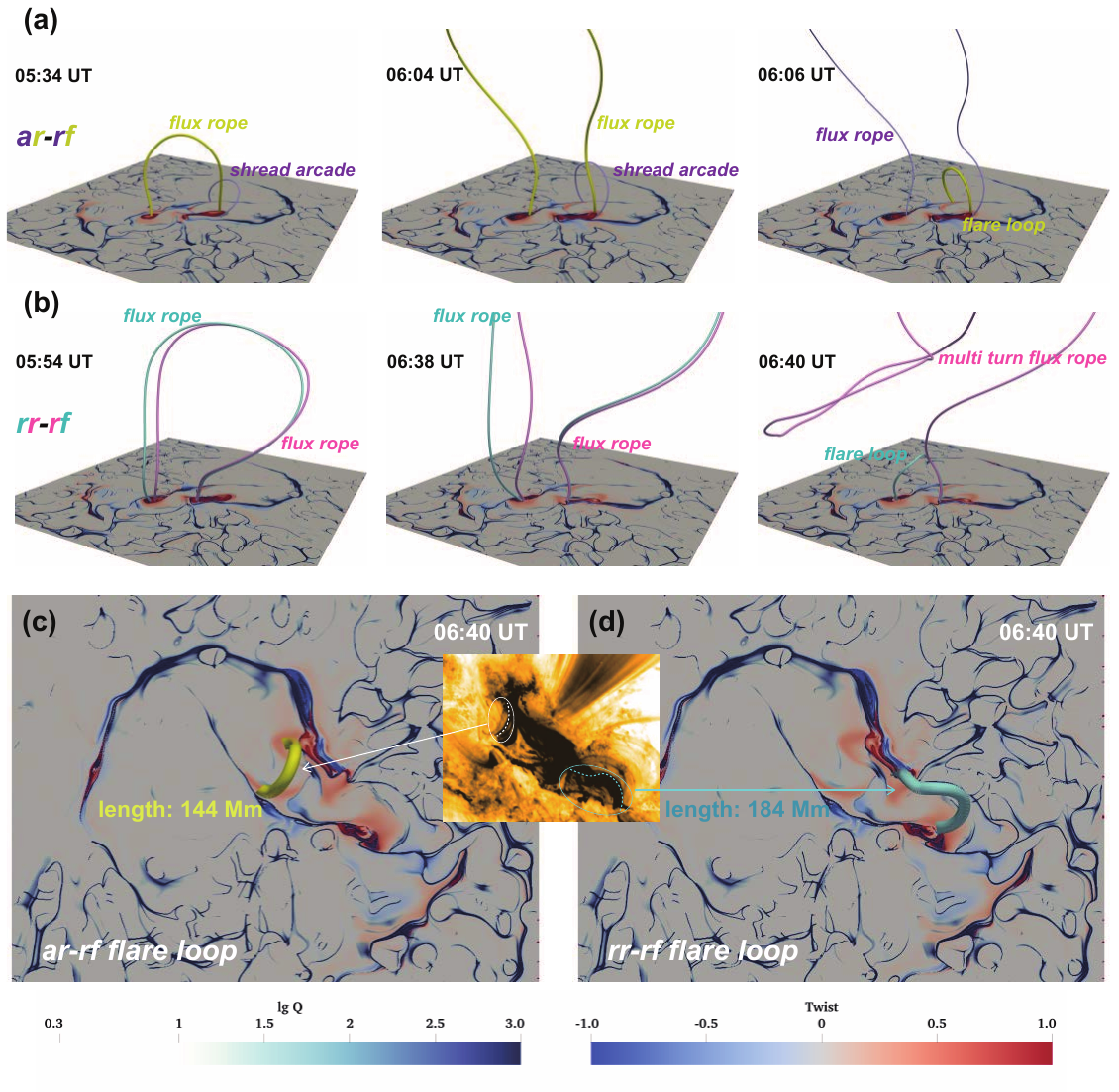}
\caption{Evolution of selected magnetic field lines under different reconnection scenarios. Each field line is labeled according to its topological connectivity at the time of plotting. The bottom planes display maps of QSL and twist, revealing double-J-shaped current ribbons. Panel (a) illustrates the $ar\text{--}rf$ reconnection process, with the resulting flare loop shown in panel (c). Panel (b) depicts the $rr\text{--}rf$ reconnection, and the newly formed flare loop is presented in panel (d). The two numerically traced flare loops are further overlaid (as dotted lines) on an AIA 171 {\AA} image plotted in the inset panel of the figure.}
\label{figure7}
\end{figure}

To investigate the reconnection process, we trace representative magnetic field lines following the approach of \citet{Aulanier2019}. Figure~\ref{figure7}(a) illustrates the geometry of the $ar\text{--}rf$ reconnection. We find that a flux-rope field line (yellow) transforms into a flare-loop field line when the separating QSLs sweep across the northern flux-rope footpoint, as outlined by the twist distribution. Conversely, a sheared-arcade field line becomes more twisted and evolves into a flux-rope field line. This exchange of field-line footpoints naturally explains the footpoint drifting in observations. Figure~\ref{figure7}(c) shows the flare loops produced by the $ar\text{--}rf$ reconnection. In addition, Figure~\ref{figure7}(b) shows the $rr\text{--}rf$ reconnection geometry in simulation. It is found that reconnection between two flux--rope field lines generates a more twisted flux-rope field line (pink) and flare loops connecting the two footpoints (cyan, also shown in Figure~\ref{figure7}(d)).

In the inset panel of Figure \ref{figure7}, we further overlay the two numerically traced flare loops on an AIA 171 {\AA} image taken at 06:40 UT. First, the location and morphology of the $ar\text{--}rf$ and $rr\text{--}rf$ flare loops in the simulation agree well with the observed flare loops in Regions 2 and 3, respectively. Second, the lengths of the $ar\text{--}rf$ and $rr\text{--}rf$ flare loops are approximately 144 Mm and 184 Mm, respectively, substantially longer than the $aa\text{--}rf$ flare loops formed at the onset of the flare ($\sim$ 17 Mm). Therefore, our data-constrained MHD simulation confirms that the flare loops generated by the $ar\text{--}rf$ and $rr\text{--}rf$ reconnection geometries are significantly longer than those produced by the $aa\text{--}rf$ reconnection, indicating that two peaks in EUV 335 {\AA} channel should be attributed to longer flare loops formed with different reconnection geometries in 3D flare model.

\begin{figure}[ht!]
\centering
\includegraphics[scale=0.8]{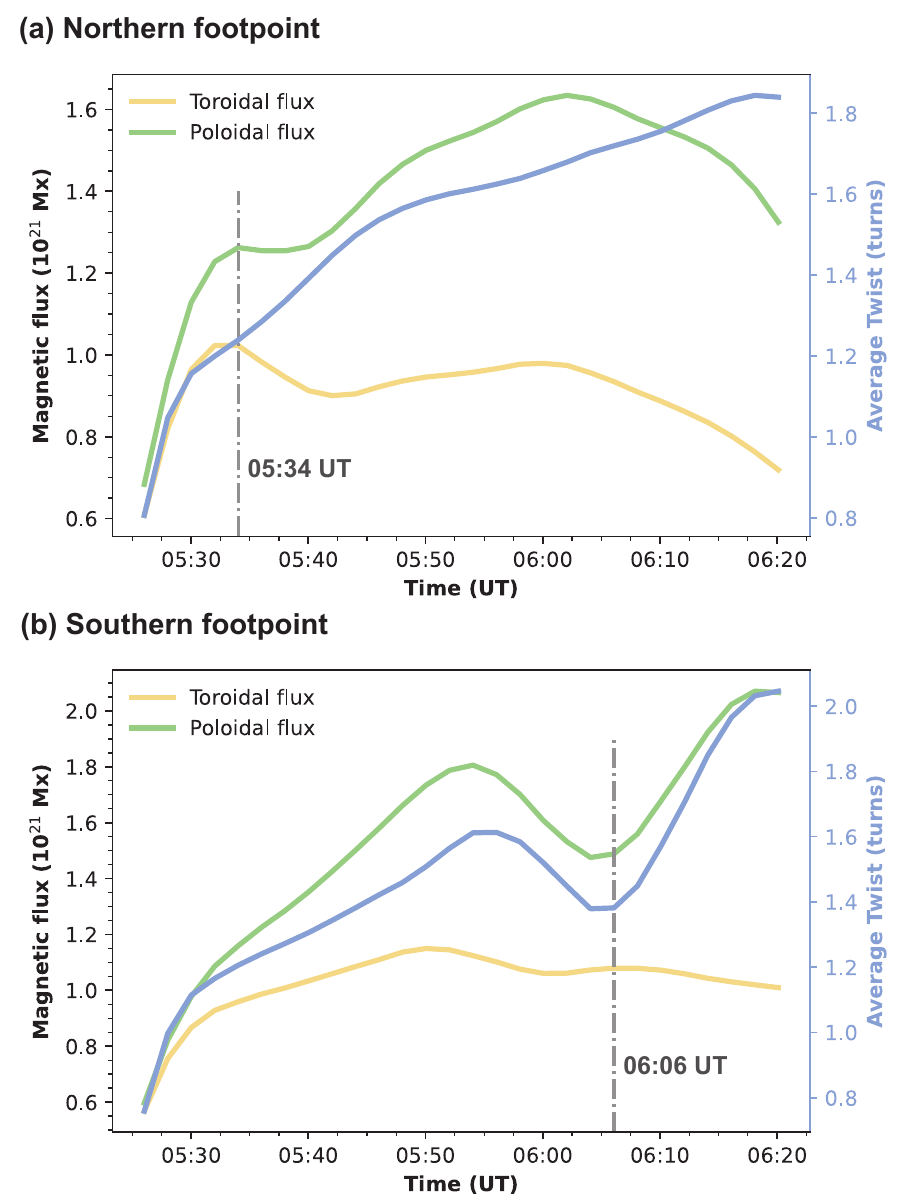}
\caption{Temporal evolution of the toroidal flux (yellow line), poloidal flux (green line), and average twist (blue line) of the flux rope. The results in panel (a) are computed based on the northern footpoint, while those in panel (b) are based on the southern footpoint. Vertical dashed lines mark the different stages of reconnection.}
\label{figure8}
\end{figure}

To further investigate the impact of 3D magnetic reconnection on the eruptive flux-rope evolution, we examine the temporal variations of the toroidal flux ($F_{t}=\int B_{z}ds$), mean twist ($\overline{T_{w}}$), and poloidal flux ($F_{p}=F_{t}\overline{T_{w}}$) for the two flux-rope footpoints (Figure~\ref{figure8}). The flux-rope footpoints are defined as regions where $\overline{T_{w}}>1$.

For the northern footpoint (Figure~\ref{figure8}(a)), both the toroidal and poloidal fluxes, as well as the mean twist, increase prior to 05:34 UT, consistent with the $aa\text{--}rf$ reconnection geometry. Afterward, the toroidal flux decreases while the poloidal flux and mean twist continue to increase, indicating a conversion of toroidal flux into poloidal flux and flare loops through the $rr\text{--}rf$ reconnection. Subsequently, both the toroidal and poloidal fluxes decrease, whereas the mean twist continues to rise. This behavior may reflect flux-rope bifurcation during the eruption. As to $ar\text{--}rf$ reconnection, for the reason pointed out above, it barely affects the evolution of the flare magnetic parameters.

For the southern footpoint (Figure~\ref{figure8}(b)), the $rr\text{--}rf$ reconnection occurs at approximately 06:06 UT, after which the toroidal flux decreases while the poloidal flux and mean twist increase. Notably, a short interval appears in which the toroidal flux, poloidal flux, and mean twist all decrease, suggesting erosion of the erupting flux rope as it interacts with the fan–spine magnetic structure in the southern domain. 

In summary, the temporal evolutions of flux and twist at the two flux-rope footpoints reflect the progression of 3D magnetic reconnection. In particular, the $rr\text{--}rf$ reconnection initiates later at the southern footpoint (06:06 UT) than at the northern one (05:34 UT), consistent with the evolution of the conjugate dimming regions shown in Figure~\ref{figure3}.

\section{Discussion and Conclusions}\label{sec:conclu}

\subsection{Overview}

Combining multi-wavelength observations with data-constrained MHD simulations, in this study, we analyze and model an M1.4 two-ribbon long duration flare that is driven by the continuous evolution of an erupting flux rope, exploring its emission characteristics in a framework of 3D magnetic reconnection geometries involved in the flux-rope eruption.

Our analysis identifies three distinct peaks in the EUV light curves (AIA 171 {\AA} and AIA 335 {\AA}; see Figure \ref{figure1}(c)--(d)), each likely corresponding to spatially distinct emission source regions. The first peak primarily originates from $aa\text{--}rf$ reconnection, the second peak from a combination of $ar\text{--}rf$ and early $rr\text{--}rf$ reconnection (Region 2), and the third peak from a subsequent, distinct $rr\text{--}rf$ reconnection (Region 3). These identifications are supported by the drifting of the footpoint (FP+) and flare ribbon (for $ar\text{--}rf$), and particularly the increase (for $aa\text{--}rf$) and subsequent decrease (for $rr\text{--}rf$) in toroidal flux at the flux-rope footpoints as recognized from conjugate dimming regions. The DEM analysis provides evidence for a multi-stage energy release process. The sequential peaking of the DEM-weighted temperature in Region 2 and Region 3, separated by approximately 6 minutes, indicates that the energy release may be governed by different magnetic reconnection processes, rather than a single impulsive event followed by gradual cooling (see Figure \ref{figure3}(f)).

Moreover, based on the approximate estimates discussed above, the cooling times of Region 2 (56 minutes) and Region 3 (76 minutes) are approximately 2.2 and 3 times that of Region 1 (25 minutes), respectively. It demonstrates that the flare loops generated in Region 2 (by $ar\text{--}rf$ and early $rr\text{--}rf$ reconnection) are longer than those from $aa\text{--}rf$ reconnection (Region 1), while the loops formed in Region 3 (by subsequent $rr\text{--}rf$ reconnection) are the longest. The different cooling times of these variously long post-flare loops determine when they reach their peak emission in the 335 \AA\ passband, thus producing distinguishable multiple peaks in the light curve. We also demonstrate that additional heating is not the cause of the multi-peak structure.

Furthermore, the data-constrained MHD simulation supports these observational findings. Based on the field-line trace technique and analysis of flux-rope magnetic flux evolution, the simulation results demonstrate different types of reconnection processes during the eruption process. The reconnection between the overlying arcade and the flux rope ($ar\text{--}rf$), combined with early $rr\text{--}rf$ reconnection, is associated with the energy release observed in Region 2. A subsequent, reconnection between flux-rope field lines themselves ($rr\text{--}rf$) in Region 3 results in the formation of a highly twisted flux rope and longer flare loops.

As a result, the observations and simulations suggest that $ar\text{--}rf$ and $rr\text{--}rf$ reconnection geometries generate longer flare loops than $aa\text{--}rf$ reconnection, and the extended cooling of these loops naturally leads to multiple EUV peaks.

\subsection{Multiple Peaks in Two-ribbon Solar Flare}

It is generally thought that multiple EUV peaks, including those observed in EUV late-phase flares, are most commonly observed in events exhibiting circular or multiple ribbons occurring within fan-spine magnetic configurations \citep{Sun2013}. First, fan-spine configurations permit several reconnection processes to occur in one flare event, e.g., null-point reconnection, CSHKP-type reconnection below the erupting flux rope, and external reconnection between the erupting flux rope and overlying fan-spine field lines \citep{Masson2009, Dai2018b}. The successive energy release from these distinct reconnection episodes can naturally generate multiple heating events and, consequently, multiple EUV peaks \citep{Liu2015}. Second, loops produced by fan-spine reconnection are typically much longer than those formed by standard CSHKP reconnection, leading to an extended cooling duration that might give rise to late-phase emission. However, \citet{Chen2020} found that approximately 40\% of EUV late-phase flares exhibit only two ribbons rather than circular or multiple-ribbon patterns. This observation raises an important question: how are multiple EUV peaks generated in these two-ribbon flare events?

The event analyzed in this study can be classified as a two-ribbon late-phase flare, in which the third EUV peak is approximately 1.13 times the amplitude of the second. We show that even a two-ribbon flare can harbor different 3D reconnection geometries, which produce flare loops of different lengths. The flare loops formed through the $ar\text{--}rf$ and $rr\text{--}rf$ reconnection geometries are at least two and three times longer, respectively, than those generated by the $aa\text{--}rf$ reconnection. Consequently, the difference in their cooling times produces the multiple EUV peaks observed in this event. It is worth noting that the 3D flare model is formulated in a simple bipolar configuration, where flare ribbons exhibit double J-shaped structures \citep{Janvier2015}. The conclusions drawn from this event may therefore help explain the origin of late-phase in two-ribbon flares.

\subsection{Predicting Flare EUV Light Curves and CME Magnetic Structures from Imaging Observations}

This work bridges the gap between imaging observations and the EUV light-curve profiles of solar flares based on 3D flare model. For example, the $ar\text{--}rf$ reconnection geometry can be identified through the drifting of flux-rope footpoints \citep{Aulanier2019, Dudik2019}, the appearance of saddle-shaped flare loops \citep{Lorincik2019}, and the lateral displacement of filament material \citep{Guo2023b}. The $rr\text{--}rf$ geometry could be associated with a decrease in magnetic flux at the flux-rope footpoint \citep{Xing2020}, the convergence of bright filament strands together with the flare loops below them \citep{Dudik2019}, and prominent supra-arcade downflows \citep{Dudik2022}. Based on the conclusions of this study, flares exhibiting these observational signatures are expected to show multiple EUV peaks, reflecting the formation of flare loops with significantly different lengths and cooling times.

Additionally, the magnetic structures of CMEs associated with these phenomena may also be more intricate. \citet{Guo2023b} showed that the $ar\text{--}rf$ reconnection geometry can cause the axial direction of a CME flux rope to deviate from that of its precursors. \citet{Guojh2024} and \citet{Guojh2025} further demonstrated that $ar\text{--}rf$ reconnection may produce a non-coherent CME flux rope without a common axis. The $rr\text{--}rf$ reconnection tends to reduce the toroidal flux and increase the twist of the resulting flux rope \citep{Aulanier2019}. These findings collectively suggest that CMEs associated with multiple-peak, two-ribbon flares may possess more twisted and intricate magnetic structures.

\begin{acknowledgments}
We are grateful to the anonymous referee for their insightful comments and suggestions. This work is supported by the National Key R\&D Program of China (2021YFA1600504, 2022YFF0503004), NSFC (12127901, 1250030413, 12333009, 12373064), China National Postdoctoral Program for Innovative Talents fellowship under Grant Number BX20240159 and Specialized Research Fund for State Key Laboratory of Solar Activity and Space Weather. The SDO data are available courtesy of NASA/SDO and the AIA and HMI science teams. The numerical calculations in this paper were performed in the cluster system of the High Performance Computing Center (HPCC) of Nanjing University.
\end{acknowledgments}

\appendix
\section{Identification of the conjugate dimmings} \label{appA}
\begin{figure}[ht!]
\centering
\includegraphics[scale=0.8]{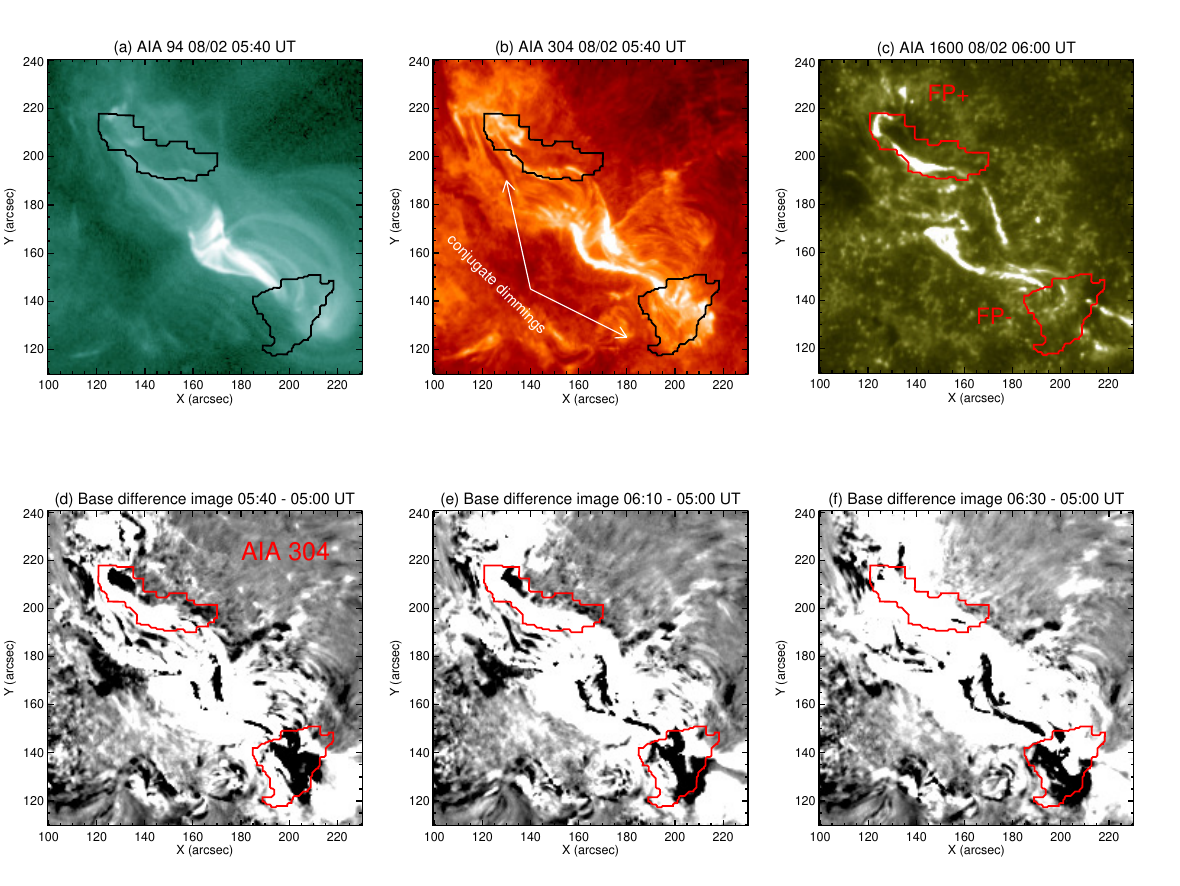}
\caption{Conjugate dimming regions observed in AIA images. Top: Snapshots in the AIA 94 \AA, 304 \AA, and 1600 \AA\ channels. ``FP+" and ``FP${}-$" indicating the positive and negative footpoints of the flux rope, respectively. Bottom: AIA 304 \AA\ base-difference image. The two contours in all panels outline the identified flux-rope footpoints.}
\label{fig_app1}
\end{figure}

Similar to \citet{Qiu2017, Wang2017, Wang2019}, we utilize conjugate dimmings to identify the flux-rope footpoints. The specific procedure for identifying valid dimming regions is as follows. For each EUV passband, potential dimming areas are first segmented from the images using a threshold-based method and then projected onto the corresponding HMI magnetogram. The identified flare ribbons (as described in \ref{sub2_2}) are also projected onto the same magnetogram. Next, only those dimming regions located near the flare ribbons are retained. The magnetic polarity of these dimming regions is calculated, and areas without a clearly dominant polarity are discarded, resulting in a set of valid dimming regions. Finally, a pixel is marked if its intensity decreases by more than 30\% compared to its average value during the quiet period and shows a sustained dimming trend before the eruption. The marked pixels from all seven EUV passbands are combined to form the two largest contiguous areas, which are taken to represent the flux-rope conjugate footpoints \citep[see][]{Wang2017, Wang2019}. The contours labeled ``FP+" and ``FP${}-$" in Figure \ref{fig_app1} mark the maximum contiguous areas of dimming pixels, identifying the flux rope's conjugate footpoints in the positive and negative magnetic fields, respectively. It can be seen that the identified edges of the conjugate dimming regions correlate well with the hook-shaped structures of the flare ribbons (especially for ``FP+").

\section{Detailed comparison between observation and simulation} \label{appB}
\begin{figure}[ht!]
\centering
\includegraphics[scale=0.4]{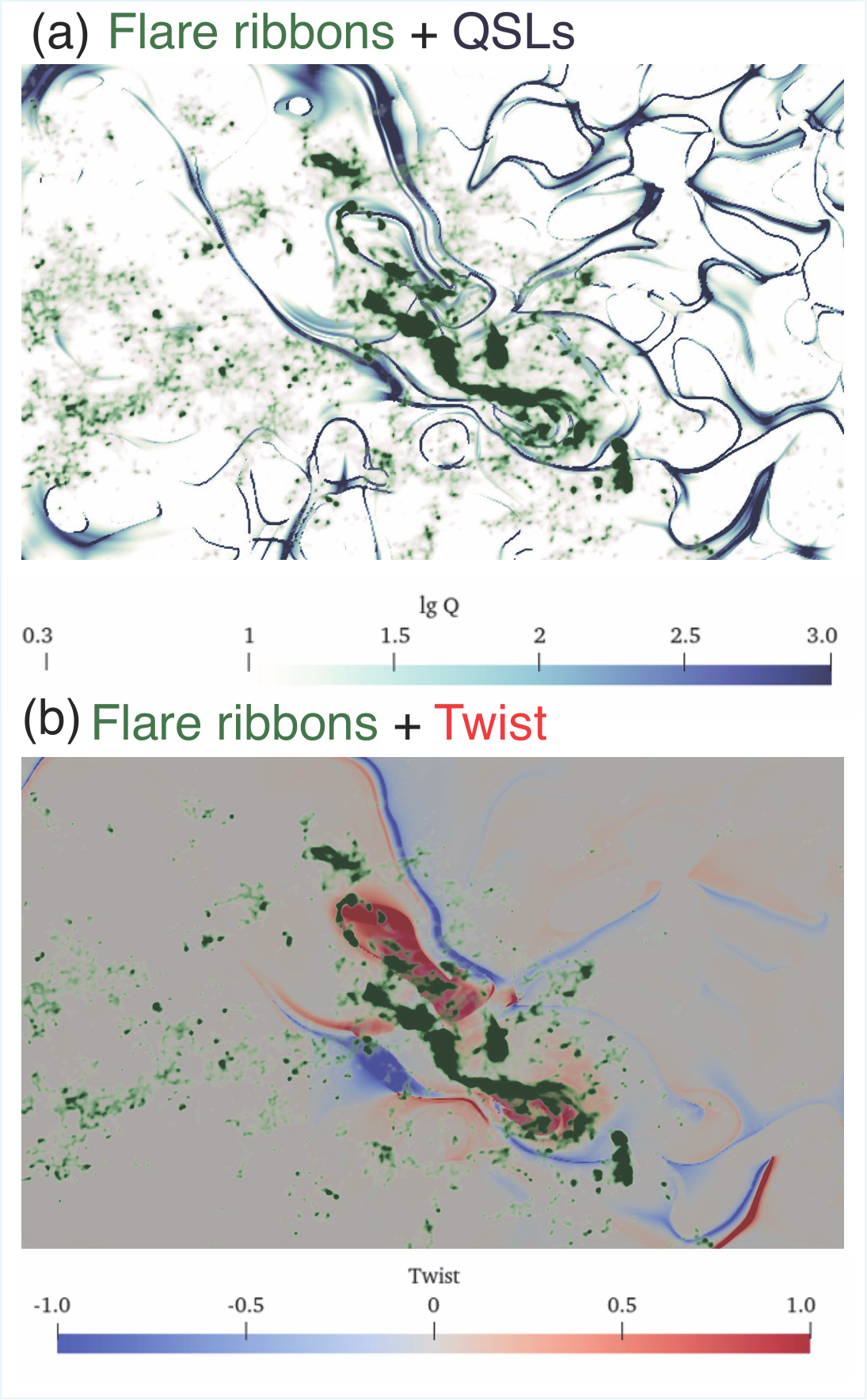}
\caption{
Comparison between AIA observations and the data-constrained simulation at the bottom boundary. Upper: AIA 1600 {\AA} image overlaid on the distribution of QSLs. Lower: The same image overlaid on the distribution of twist. The Q and twist maps are the same as in Figure \ref{figure5} at the corresponding time (05:44 UT).}
\label{fig_app2}
\end{figure}

In Figure \ref{fig_app2} we overlay the AIA 1600 {\AA} image on the simulated distributions of QSL and twist (the second row of Figure \ref{figure5}). It is seen that the hook-shaped flare ribbons (green) show a spatial alignment with both the Q-map and the regions of high twist. Such a direct comparison corroborates the robustness of our numerical simulation in reproducing the observed flare features.

\bibliography{ms}{}
\bibliographystyle{aasjournal}

\end{CJK*}
\end{document}